\documentclass{elsart}
\usepackage{graphicx,amssymb}
\journal{New Astronomy}

\begin{document}
\begin{frontmatter}

\title{The Chamomile Scheme: An Optimized Algorithm for $N$-body simulations on Programmable Graphics Processing Units}
\author{Tsuyoshi Hamada},
\ead{thamada@riken.jp}
\ead[url]{http://progrape.jp/cs/}
\author{Toshiaki Iitaka}
\ead{tiitaka@riken.jp}
\ead[url]{http://www.iitaka.org/}

\address{Computational Astrophysics Laboratory, RIKEN, 2-1 Hirosawa, Wako, Saitama 351-0198, Japan}

\begin{abstract}
We present an algorithm named ``Chamomile Scheme''. The scheme is
fully optimized for calculating gravitational interactions on the
latest programmable Graphics Processing Unit (GPU), NVIDIA
GeForce8800GTX, which has (a) small but fast shared memories (16 K
Bytes $\times$ 16) with no broadcasting mechanism and (b) floating
point arithmetic hardware of 500 Gflop/s but only for single
precision. Based on this scheme, we have developed a library for
gravitational $N$-body simulations, ``CUNBODY-1'', whose measured
performance reaches to 173 Gflop/s for 2048 particles and 256 Gflop/s
for 131072 particles.
\end{abstract}
\end{frontmatter}

\section{Introduction}

\subsection{The $N$-body simulation}

Astronomical $N$-body simulations have been widely used to
investigate the formation and evolution of various astronomical
systems, such as planetary systems, globular clusters, galaxies,
clusters of galaxies, and large scale structures of the universe.  
In such simulations, we treat planetesimals, stars, or galaxies as
particles interacting with each other. We numerically evaluate
interactions between the particles and advance the particles according
to the Newton's equation of motion.

In many cases, the size of an astrophysical $N$-body simulation is
limited by the available computational resources.  Simulation of pure
gravitational $N$-body system is a typical example.  Since the gravity
is a long-range interaction, the calculation cost for interactions
between all particles is $O(N^2)$ per
time-step for the simplest scheme, where $N$ is the number of particles
in the system.  We can reduce this $O(N^2)$ calculation cost to
$O(N\log N)$, by using some approximated algorithms, such as the
Barnes-Hut tree algorithm\cite{1986Natur.324..446B}, but the scaling
coefficient is pretty large.  Thus, the calculation of the interaction
between particles is usually the most expensive part of the entire
calculation, and thus limits the number of particles we can handle.
Smoothed Particle Hydrodynamics
(SPH)\cite{1977AJ.....82.1013L}\cite{Gingold_1977}, in which particles
are used to represent the fluid, is another example.  In SPH
calculations, hydro-dynamical equation is expressed by short-range
interaction between particles.  The calculation cost of this SPH
interaction is rather high, because the average number of particles
which interact with one particle is fairly large, typically around 50,
and the calculation of single pairwise interaction is quite a bit more
complex compared to gravitational interaction.

In astrophysical $N$-body simulations,
the most important interaction is gravitational force.
Using computer resources, we calculate the gravitational force 
of an $i$-th particle from many $j$-th particles in the
following equation:
\begin{equation}
\mathbf{ a}_i = \sum_j {m_j \mathbf{ r}_{ij} \over (r_{ij}^2 +
\varepsilon^2)^{3/2}}
\label{eq:aij}
\end{equation}
where $\mathbf{ a}_i$ is the gravitational acceleration of $i$-th
particle (here after, we call as $i$-particle), $\mathbf{ r}_j$ and
$m_j$ are the position and mass of $j$-th particle(here after
$j$-particle) respectively, and $\mathbf{ r}_{ij} = \mathbf{ r}_{j} - \mathbf{
r}_{i}$.

Astrophysics is not the only field where the $N$-body simulation is
used.  Molecular dynamics (MD)\cite{MD_Book} simulation and boundary
element method (BEM) are examples of numerical methods where each
element of the system in principle interacts with all other elements
in the system.  In both cases, approaches similar to the Barnes-Hut
tree algorithm or FMM\cite{1987JCoPh..73..325G} help to reduce the
calculation cost, but the interaction calculation still dominates the
total calculation cost.

One extreme approach to accelerate the $N$-body simulation is to
build a special-purpose computer for the interaction calculation.  Two
characteristics of the interaction calculation make it well suited for
such approach.  Firstly, the calculation of pairwise interaction is
relatively simple.  In the case of gravitational interaction, the total
number of floating-point operations (counting all operations, including
square root and divide operations) is only around 20.  So it is not
inconceivable to design a fully pipelined, hard-wired processor dedicated
to the calculation of gravitational interaction.  For other application
like SPH or molecular dynamics, the interaction calculation is more
complicated, but still hardware approach is feasible.  Secondly, the
interaction is in its simplest form all-to-all.  In other words, each
particle interacts with all other particles in the system.  Thus, there
is lots of parallelism available.  In particular, it is possible to
design a hardware so that it calculate the force from one particle to
many other particles in parallel.  In this way we can reduce the
required memory bandwidth.  Of course, if the interaction is of
short-range nature, one can implement some clever way to reduce
calculation cost from $O(N^2)$ to $O(N)$, and the reduction in the
memory bandwidth is not as effective as in the case of true $O(N^2)$
calculation.

The approach to develop specialized hardware for gravitational
interaction, materialized in the GRAPE ("GRAvity piPE") project
\cite{1990Natur.345...33S}\cite{Makino_1998}, has been fairly
successful, achieving the speed comparable or faster than the fastest
general-purpose computers. A major limitation of GRAPE is that it
can't handle anything other than the interaction through $1/r$
potential. It is certainly possible to build a hardware that can
handle arbitrary central force, so that molecular dynamics calculation
can also be
handled\cite{1993PASJ...45..339I}\cite{1996ApJ...468...51F}
\cite{1999MolSim.21..201N}\cite{2003SC..Taiji}\cite{2006SC...Narumi}.

However, to design a hardware that can calculate both the
gravitational interaction and, for example, an SPH interaction is
quite difficult. Actually, to develop the pipeline processor just for
SPH interaction turned out to be a rather difficult task
\cite{1999AG..Spurzem}\cite{1999numa.conf..429Y}\cite{1999...kuberka}\cite{LKM02}\cite{2002JSPS..Spurzem}.
This is provably because the SPH interaction is much more complex than
gravity.

Instead of design a hardware, we can use a programmable hardware to
accelerate many kinds of interaction.  The Graphics Processing
Unit(GPU) is one of such programmable hardware.

\subsection{The Graphics Processing Unit}

The history of Graphics Processing Unit (GPU) is
important to understand its possibility.

In early times, GPU was
a special-purpose device for processing graphics. The GPU implemented
numbers of graphics pipelines in a way that makes running them much
faster than drawing directly to the screen with the conventional CPU.
For decades, the GPU basically had been very similar to GRAPE, which
had many hard-wired pipelines strictly dedicating to graphics processing
such as drawing rectangles, triangles, circles, and arcs on the
screen.

Recently, the GPU had been very efficient in calculating and
displaying computer graphics with their highly parallel structure.
Corresponding with the advent of graphics
libraries(such as the 8th-generation DirectX, or OpenGL), GPU
vendors began to add some programmable hardware instead of the
hard-wired pipeline.

In this generation of GPU, such programmable hardware was called the
programmable shader.  After the GPU implemented the programmable
shader, the generality of GPU had improved a little.  Using the
programmable shader, graphics programmer was able to control GPUs not
only in polygon level but also in pixel level through their program
code.

As the shading calculation has much compute-intensive and highly parallel
characteristic, GPU vendors had to implement large numbers of
arithmetic hardwares onto their GPUs. As a result, the performance
improvement of GPU had surpassed that of CPU.

\subsection{The GPU-based Computing}

As the GPU had general-purpose functionality and the performance of
GPU were improving rapidly, some people thought they wanted to use the
GPU not only for graphics processing but for numerical computing such
as physics, chemistry or astrophysics simulations.  Such demands
pierced and moved GPU vendors to open methods for people in numerical
computing, and GPU vendors began to supply methods for accessing
general-purpose functions on GPU for numerical computing.  People
began to call such method as the General-Purpose Computation on
GPUs(GPGPU)\cite{2005SIGGRAPH..Harris} or GPU-based computing.

In $N$-body simulations, the GPU-based computing also could offer the
level of flexibility that was impossible to achieve with the
conventional GRAPE approaches. GPU is a mass-produced hardware,
consisting of a large number of floating-point hardwares in a SIMD(Single
Instruction stream, Multiple Data stream) fashion. By programming
these floating-point hardwares, we can implement an arbitrary pipeline
processor on GPU. Thus, a single hardware can be used to implement
various interactions, such as that for gravity, SPH, and others.

Recently the approach to accelerate $N$-body simulations with GPUs
has become an active area of research.  For
molecular dynamics simulation, Nyland et al. reported 4096 body
simulation with $290 \times 10^6$ (pair/second) $\times 38$
(flop/pair) $\simeq$ $11$ (Gflop/s)\cite{2004GPGPU..Nyland}.  Mark
Harris reported $N$-body simulation of 8192 body system with $480
\times 10^6$ (pair/second) $\times 38$ (flop/pair) $\simeq 18$
(Gflop/s)\cite{2005Game..Harris}, and $672 \times 10^6$ (pair/second)
$\times 38$ (flop/pair) $\simeq 26$
(Gflop/s)\cite{2005SIGGRAPH..Harris}.

There were some more requirements needed for applying the conventional
GPUs to $N$-body simulations. The most important requirement was about
the performance of the conventional GPUs.  Compared with that of the
conventional GRAPE, the GPUs used by Harris and Nyland had the peak
performance of only around 100 Gflop/s.  On the other hand, in the
same age, the GRAPE-6Af\cite{2005astro-ph/0504407} in a PCI add-in
fashion already had the peak performance of 131.3 Gflop/s. According
to these performance differences, people in astrophysics did not try
to use GPUs as their computing resources.

\subsection{The Compute Unified Device Architecture (CUDA)}

There also existed another requirement for the conventional GPU.
It was the programming method for GPU, which remained quite tricky.

One of widely used methods for programming GPU is the Cg language.
The Cg language\footnote{{\tt http://www.nvidia.com/}} was developed
by the NVIDIA corporation.  Using Cg, we can use 
both of NVIDIA's and ATI's GPUs as computing devices.  However
we should access the GPUs only through the graphics libraries, such as
OpenGL or DirectX, in the Cg language.  Such programming methods are often
quite tricky for people in astrophysics.

Naturally, the GPU vendors had known such circumstances.  In November
2006, the NVIDIA corporation opened a quite smart programming method
for their GPUs, the Compute Unified Device Architecture (CUDA).  Using
the CUDA, we can write the GPU's program as if we use the standard C
language and we don't need to suffer with any graphics libraries.

\subsection{The GeForce8800GTX system}

Returning our explanation to about the peak performance of GPU, 
the situation has been changed dramatically in these several months.

At the middle of November 2006, the NVIDIA corporation has released
a modern GPU, NVIDIA GeForce8800GTX. Its peak performance reaches even
500 Gflop/s.

\begin{figure}
\begin{center}
\includegraphics[width=1.0\textwidth]{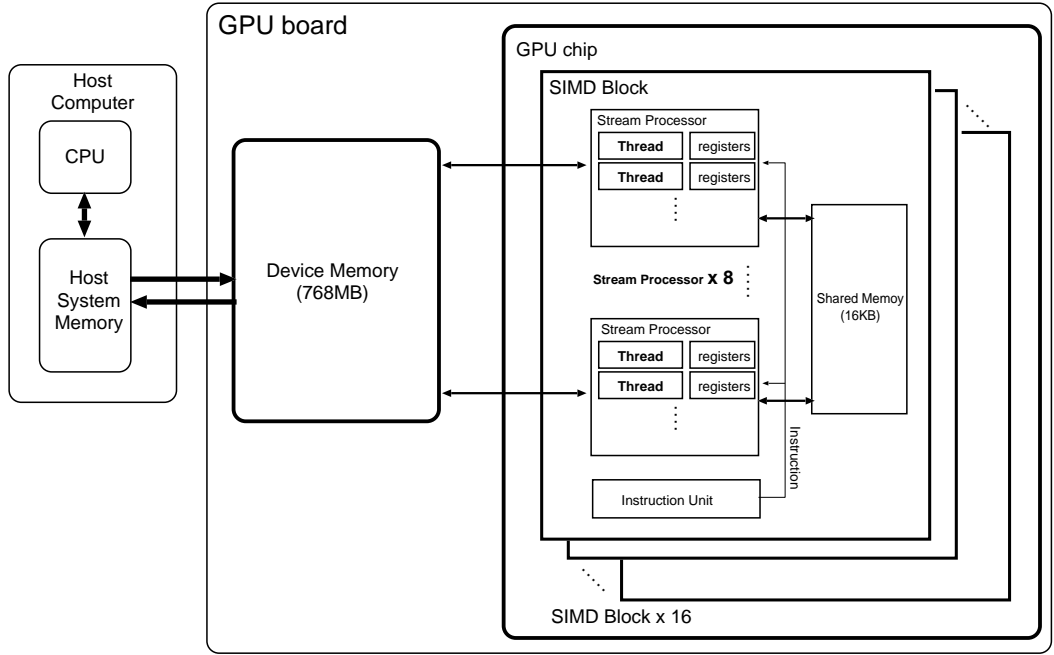}
\end{center}
\caption{The basic structure of GeForce8800GTX system.}
\label{fig:G80}
\end{figure}

Figure. \ref{fig:G80} shows the basic structure of GeForce8800GTX
system\footnote{\tt http://www.nvidia.com/} It is connected to a host
computer, which works as an attached processor.  The GeForce8800GTX
system is composed of one GPU chip and device memory
of 768 M Bytes GDDR3-DRAM(Graphics Double Data Rate 3 Dynamic Random
Access Memory).  The GPU chip consists of 128
processors(Stream Processors in Figure.\ref{fig:G80}). 

The stream processor is a general-purpose processor, which has three
floating-point units(floating-point multiply, floating-point multiply
and addition). Each stream processor works at 1.35 GHz. Then the peak
performance of the GeForce8800GTX equals 518.4 Gflop/s (128
[processors] $\times$ 3 [float/processor] $\times$ 1.35 [GHz]).

The detailed features of the GPU chip in the GeForce8800GTX system are
as follows.  All of the processors are divided into 16 blocks(SIMD
Blocks in Figure.\ref{fig:G80}).  Each block has 8 processors in a
SIMD fashion and a quite fast but small shared memory(16 K Bytes
each). The 8 processors in a SIMD block can work with a single
instruction, and they can synchronize and communicate with each other
through the shared memory.  On the other hand, the processors in
different SIMD blocks can not synchronize with each other. Each
processor can execute many tasks with time-multiplexing as many as the
processor can allocate its registers to different tasks.

\subsection{Glanced difference between the GPU and the GRAPE}

Then, what is the fundamental difference between the modern GPU and the
conventional GRAPE in $N$-body simulations ?

\begin{figure}
\begin{center}
\includegraphics[width=1.0\textwidth]{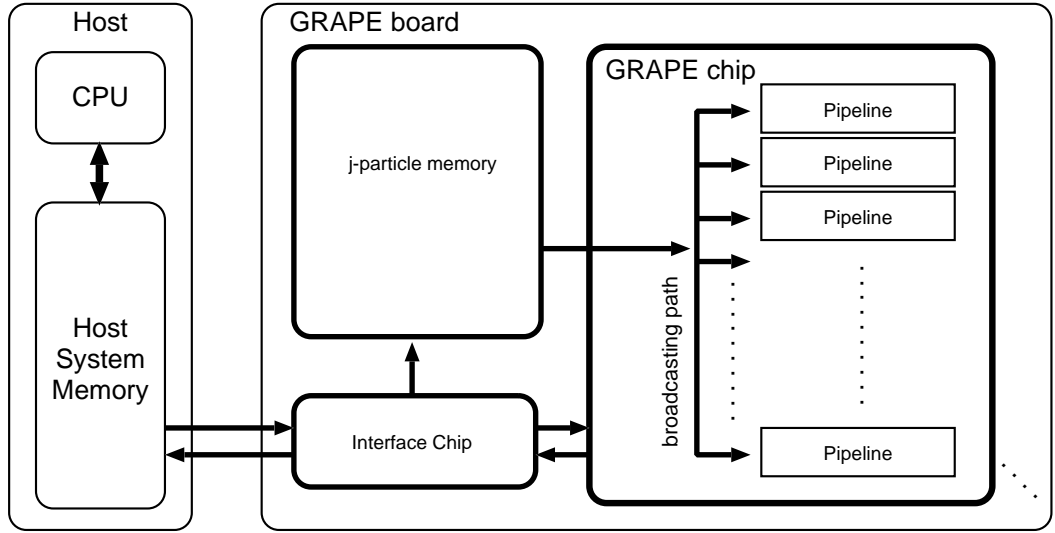}
\end{center}
\caption{The basic structure of GRAPE system}
\label{fig:GRAPE}
\end{figure}

Figure.\ref{fig:GRAPE} shows the basic structure of GRAPE system.
The GRAPE implements many pipelines with hard-wired arithmetic units on a chip
to calculate the interactions between particles in the equation(\ref{eq:aij}).
All pipelines on GRAPE chips are connected directly with large amount of 
memory(8 M Bytes of SSRAMs per chip for GRAPE-6) and have a
broadcasting mechanism. The path between all pipelines and memory
hang in the same line and the data elements of $j$-particle
can be broadcasted to all pipelines for every clock cycles.
Then, the pipelines can calculate the gravitational interactions clock
by clock and the time spent for the data moving
can be hidden behind the pipeline computations.

Another feature of the GRAPE is that the pipeline implements
accumulation units for high precision.  The pipeline of GRAPE
implements accumulation unit with 64-bit accuracy.  In the case of
accumulating force exerted from large number of particles, it is quite
important because the subtraction between two values with similar size
causes a very big error (cancellation of significant digits) and the
error accumulates when repeated many times(Here after, we call such 
error in accumulation as the cumulative error).

For $N$-body simulations,
the big memory with broadcasting mechanism and the high precision
arithmetic hardware for accumulation are the most important features of
the traditional GRAPE.  On the other hand, small
but fast shared memories(16 K Bytes $\times$ 16) with no broadcasting
mechanism and floating point arithmetic hardware of 500 Gflop/s but
only for single precision are the most important features of the
modern GPU.

Then, what everyone wants to know is the possibility of the modern GPU in $N$-body simulations. 
Is the modern GPU suitable for $N$-body simulations ?

Our conclusion described in this paper is "almost yes", but a lot of
discussions should be done by many people in astrophysics.  And it is
glad that our scheme named ``Chamomile Scheme'' becomes the first one
step for future discussions.


The plan of this paper is as follows.
In section 2, we describe the Chamomile Scheme.
In section 3, we demonstrate the performance of the modern GPU by our
library, CUNBODY-1.
And the section 4 is for discussions.

\section{The Chamomile Scheme}

\subsection{Overview of the Chamomile Scheme}

In this section, we describe our Chamomile Scheme (CS),
which is fully optimized for calculating gravitational interactions
on the modern GPU, NVIDIA GeForce8800GTX.
The advantage of the CS is that we can use the
features of the GPU:
\begin{enumerate}
\item[(1)] small but fast shared memories with no broadcasting mechanism,
\item[(2)] single precision floating point arithmetic hardware but ultra high-speed,
\end{enumerate}
effectively.

\begin{figure}
\begin{center}
\includegraphics[width=1.0\textwidth]{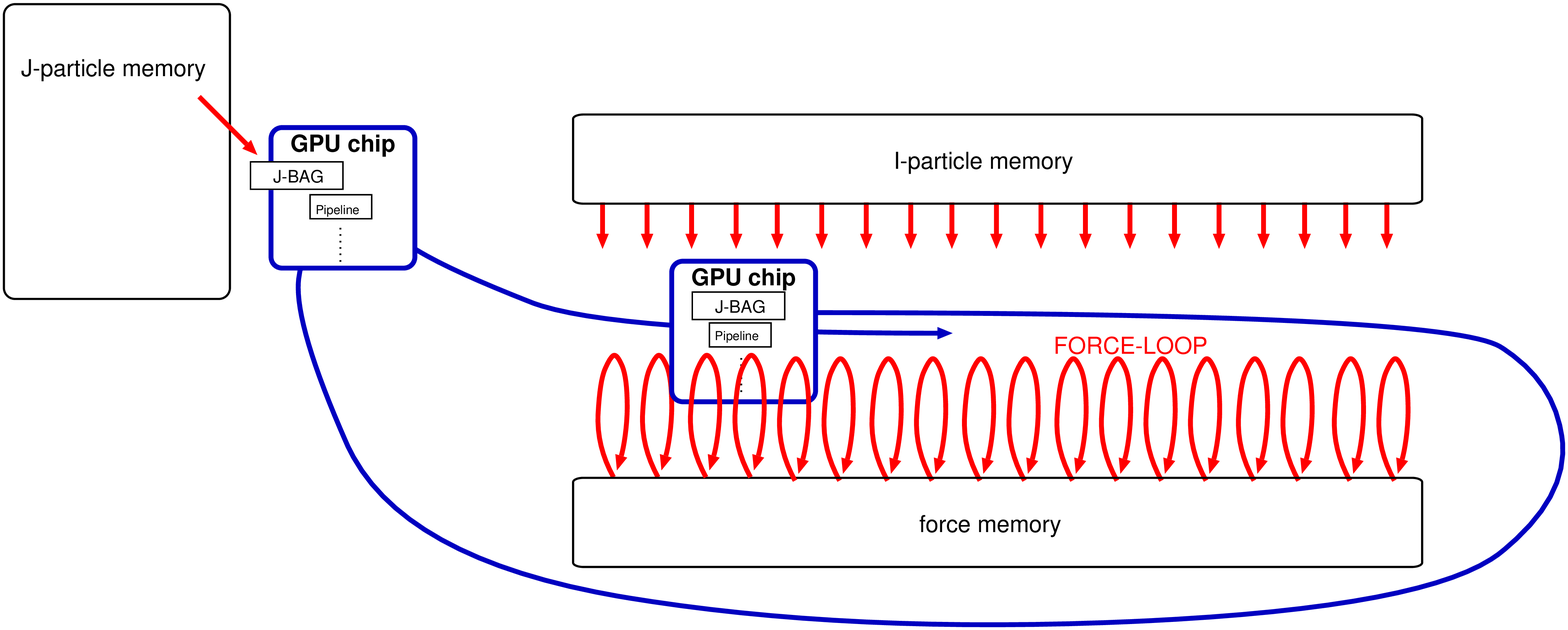}
\end{center}
\caption{Overview of the Chamomile Scheme.}
\label{fig:CS}
\end{figure}

\begin{figure}
\begin{center}
\includegraphics[width=0.9\textwidth]{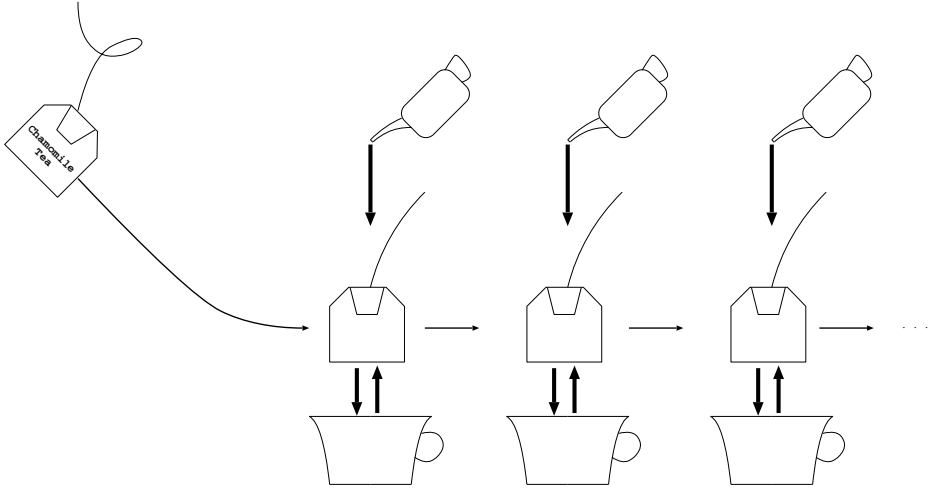}
\end{center}
\caption{An image we are inspired by}
\label{fig:inspiration}
\end{figure}

Firstly, we show the overview of our CS in Figure.\ref{fig:CS}.
The blue locus expresses the
behavior of a GPU chip, and the red lines represents the data transfers
between the GPU chip and the device memory. 
The device memory are divided into three regions: 
a region for all $j$-particles(here after, $j$-particle memory),
a region for all $i$-particles(here after, $i$-particle memory),
and a region for all force(here after, force memory).

At first, the GPU chip takes segmental $j$-particles(here after,
J-BAG) into its shared memory.  After taking the J-BAG, the GPU runs
through between $i$-particle memory and force memory, and come back to
$j$-particle memory again. This GPU chip's behavior expressed by the blue locus
is repeated again and again until all J-BAGs are taken from
$j$-particle memory.

While the GPU chip are running, the J-BAG interacts with all
$i$-particles and the force calculated by this interaction are stored
into force memory again and again(here after, FORCE-LOOP).

Our idea of CS comes from such behavior of GPU chip.  When we were
taking chamomile tea, suddenly we got an image of such GPU's behavior
illustrated in Figure. \ref{fig:inspiration}.  So we have named our
algorithm as the ``Chamomile Scheme''.

\subsection{Details of the Chamomile Scheme}

\begin{figure}
\begin{center}
\includegraphics[width=1.0\textwidth]{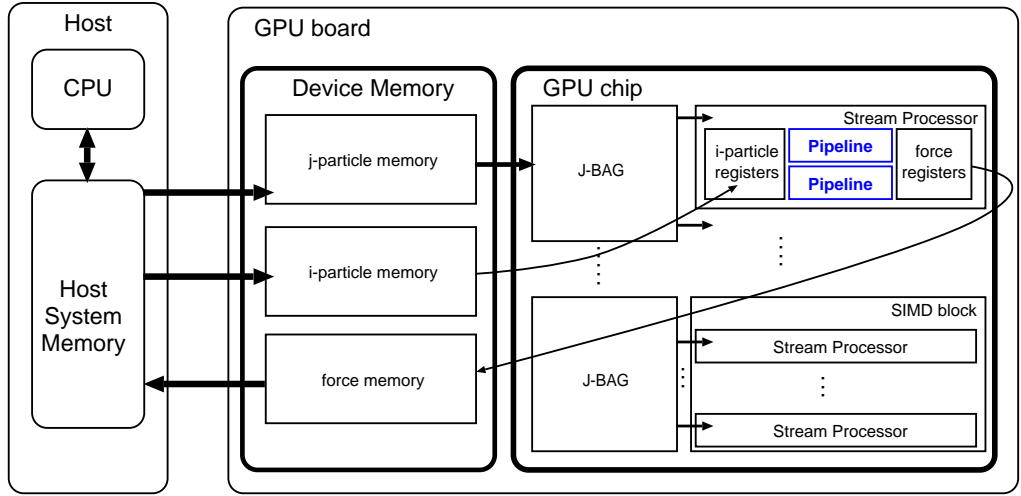}
\end{center}
\caption{The detailed implementation of the Chamomile Scheme on the GeForce8800GTX}
\label{fig:GPUstruct}
\end{figure}

Figure \ref{fig:GPUstruct} shows the detailed implementation of the CS
on the GeForce8800GTX system.  The device memory implements
$i$-particle memory, $j$-particle memory and force memory.  The shared
memory on each SIMD block implements the J-BAG for segmental
$j$-particles. The stream processor executes just as if many pipelines
are implemented and work simultaneously.  The stream processor
implements registers for $i$-particles(here after, $i$-particle
registers) and force (here after, force registers). The force
registers are divided into two regions, backup force registers and
current force registers.

Based on the CS, calculation proceeds in the following steps:
\begin{enumerate}
\item[(a)] The Host computer sends all of $i$- and $j$-particles to $i$-particle memory and $j$-particle memory, respectively.
\item[(b)] The GPU divides $j$-particles into J-BAGs with the size of shared memory.
\item[(c)] The GPU divides $i$-particles into blocks(here after, I-POTs) with the size of all pipelines.
\item[(d)] All pipelines gather the J-BAG from $j$-particle memory to their shared memory.  (J-PACKING)
        \begin{enumerate}
        \item[(e)] All pipelines gather $i$-particles from $i$-particle memory to $i$-particle registers.
        \item[(f)] All pipelines gather force calculated previously to the backup force registers.
        \item[(g)] All pipelines clear  the current force registers by zero.
        \item[(h)] All pipelines calculate interactions between the I-POT and the J-BAG while summing up the current force registers.
        \item[(i)] All pipelines add their current and backup force registers, and scatter the force to force memory.
        \item[(j)] GPU repeats (e) $\sim$ (j) for all I-POTs.
        \end{enumerate}
\item[(k)] The GPU repeats (d) $\sim$ (k) for all J-BAGs.
\item[(l)] The GPU sends back all of force to the host computer.
\end{enumerate}

Here, we call the process of (d) as J-PACKING, and (f) $\sim$ (i) as FORCE-LOOP.

Using J-PACKING, we can reduce the transfer time for $j$-particles,
because the shared memory is much faster than the device memory.  For the
GeForce8800GTX system, the latency of the shared memory is about 2
clock cycles. On the other hand, the latency of the device memory is
about 200 $\sim$ 300 clock cycles.

\begin{figure}
\begin{center}
\includegraphics[width=0.5\textwidth]{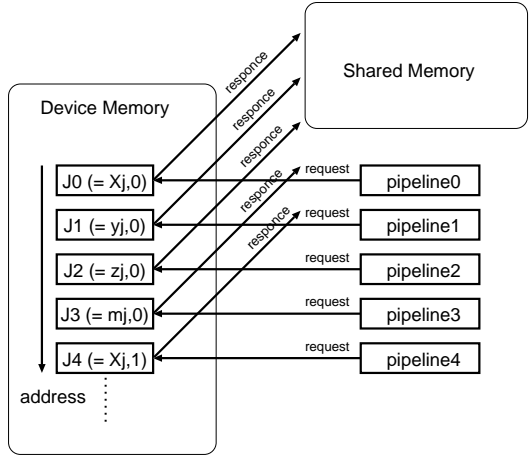}
\end{center}
\caption{The sequential accessing for the J-PACKING.}
\label{fig:jblock-tech}
\end{figure}

The J-PACKING in process (d) needs additional
time compared with the case without the shared memory.
For every J-PACKING, we need to shift the $j$-particles from the device memory
to the shared memory.

However the time spent for the J-PACKING doesn't become a bottleneck.
This is because the J-PACKING occurs only once while the GPU are
running through between $i$-particle memory and force memory.

Though the J-PACKING is not a bottleneck, there is a slight technique
for advancing the bandwidth of J-PACKING.
Figure.\ref{fig:jblock-tech} shows the technique.  
The device memory consists of GDDR3-DRAMs, by which we can use a burst accessing.
While building the J-BAG, each pipeline sends each request and the device memory
outputs $j$-particles. If the requests are sequential, the outputs also become sequential and
the bandwidth of J-PACKING improves.

\subsubsection{The Virtual Multiple Pipelining}

\begin{figure}
\begin{center}
\includegraphics[width=0.5\textwidth]{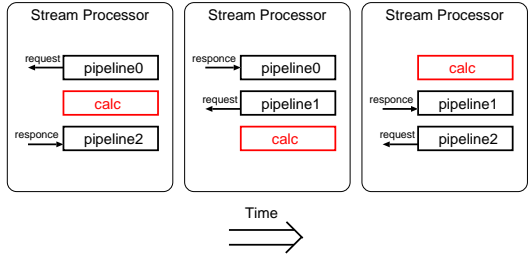}
\end{center}
\caption{An example of 3-way virtual multiple pipelines(VMPs) for a Stream Processor in
the CS algorithm. In fact, we use 16-way VMPs for our implementation.}
\label{fig:vpipe}
\end{figure}

To improve the bandwidth of the J-BAG more, we can use the NVIDIA's
multi-threading function on the stream processor.  The J-BAG doesn't
have no broadcasting mechanism while sending $j$-particles to
pipelines.  Using the multi-threading, we can hide the time for
gathering $j$-particles behind the arithmetic operations on pipelines.
Here after, we call such the usage of the multi-threading as virtual
multiple pipelining and the multi-threaded pipelines as the virtual
multiple pipelines(VMPs).

The original idea of VMP comes from
Makino et al.(1997)\cite{1997ApJ...480..432M}. The only difference
between Makino et al. and the CS is just the substance
actually working. For Makino et al., the substance is the hard-wired
pipeline and for the CS, it is the stream processor.

Figure.\ref{fig:vpipe} illustrates the VMPs on a stream processor.  In
this figure, the time goes by left to right in each clock cycle.  At
the first clock cycle, the first pipeline(pipeline0 in
Figure. \ref{fig:vpipe}) requests a $j$-particle to the J-BAG. At the
same time, the second pipeline(pipeline1) executes an arithmetic operation, and the
last pipeline(pipeline2) receives
$j$-particle from the J-BAG simultaneously.
At the next clock cycle, each operation shift to the next pipeline.
The last stage is also the same manner.
So, applying the virtual multiple pipelining, 
the time for gathering $j$-particles can be hidden by arithmetic operations
on the stream processor.

Up to now, we can understand that the CS uses one of the feature of GPU:
(1) small but fast shared memories with no broadcasting mechanism,
effectively.

\subsubsection{The FORCE-LOOP}

Secondly, we explain how the CS uses another feature of the
GPU: (2) single precision floating point arithmetic hardware but ultra
high-speed, effectively.

\begin{figure}
\begin{center}
\includegraphics[width=0.5\textwidth]{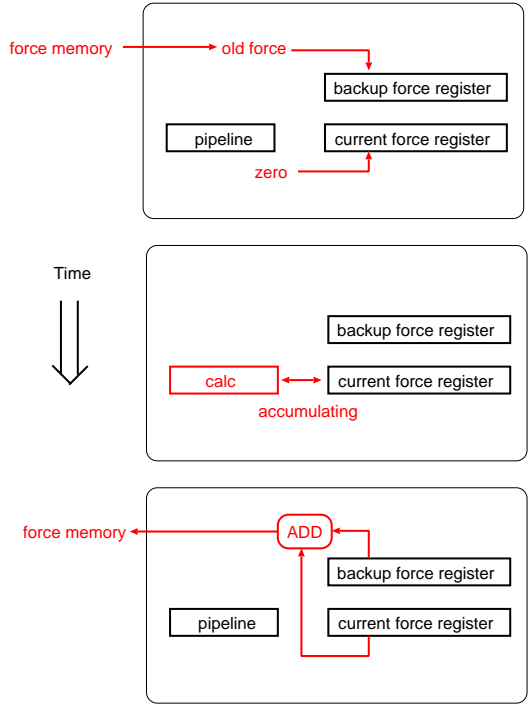}
\end{center}
\caption{The FORCE-LOOP.}
\label{fig:force-loop}
\end{figure}

We named the process between (f) $\sim$ (i) as the FORCE-LOOP in
previous subsection. 
Figure.\ref{fig:force-loop} shows the steps of the FORCE-LOOP.  In the
first stage, each pipeline gathers force (old force in this figure)
previously calculated from the force memory to the backup force
registers. Simultaneously, the pipeline clears the current force
register by zero.  At the next stage, the pipeline calculates
interactions between J-BAG and I-POT. While the calculation, the
pipeline is summing up the current force register.  After the
calculation is finished, the pipeline adds the current force register
and the backup force register, and gets back the force added to the
force memory.

In this manner, the calculation of interaction for
equation(\ref{eq:aij}) can be reformed to the following equation:
\begin{equation}
\mathbf{ a}_i =
  (...((\sum_{j=0}^{n_{jb}-1} {m_j \mathbf{ r}_{ij} \over (r_{ij}^2 +\varepsilon^2)^{3/2}})
       + \sum_{j=n_{jb}}^{2 n_{jb}-1} {m_j \mathbf{ r}_{ij} \over (r_{ij}^2 +\varepsilon^2)^{3/2}} )
   ...) + \sum_{j=nj-n_{jb}}^{n_j-1} {m_j \mathbf{ r}_{ij} \over (r_{ij}^2 +\varepsilon^2)^{3/2}}, 
\label{eq:force-loop}
\end{equation}
where $n_j$ is the number of all $j$-particles stored in $j$-particle
memory.  $n_{jb}$ is the number of $j$-particles in a J-BAG.  
By this reformation, the cumulative error strongly depends on the total
number of J-BAG, ($\frac{nj}{n_{jb}}$). 
For GeForce8800GTX, the size of the shared
memory is 16 K Byte and $\frac{nj}{n_{jb}}$ becomes around 4(for 2048
particles) $\sim$ 256(for 131072 particles) at most.
For this reason, we can reduce the cumulative error by the FORCE-LOOP.

\section{CUNBODY-1 and the Performance}

Based on the Chamomile Scheme, we have developed a library for
gravitational $N$-body simulations, CUNBODY-1 (CUDA N-BODY version 1).
And in this section, we describe the performance results of the CUNBODY-1.

\subsection{The Library Structure}

\begin{figure}
\large
\begin{verbatim}
void force(double xj[][3],
           double mj[],
           double xi[][3],
           double eps2,
           double a[][3],
           int ni,
           int nj);
\end{verbatim}
\caption{The structure of subroutine of our library}
\label{fig:libforce}
\end{figure}

Figure.\ref{fig:libforce} shows the subroutine structure of the CUNBODY-1.
\verb|ni| and \verb|nj| are the total number of $i$-particles and
$j$-particles, respectively.
\verb|xj| represents position of $j$-particles (three dimensions), 
\verb|mj| represents mass of $j$-particles,
\verb|xi| represents position of $i$-particles (three dimensions),
\verb|eps| is a softening parameter,
and 
\verb|a| represents force calculated by a GPU.
The CUNBODY-1 library can be used from standard C/C++/Fortran languages, because
we use the CUDA library to implement the CUNBODY-1.
We can see the CUNBODY-1 library interface is simple and available not only
for the direct integration algorithms but also for other clever algorithms such as 
the Barnes-Hut tree algorithm\cite{1986Natur.324..446B}, etc.

\begin{figure}
\scriptsize
\begin{verbatim}
int ni, nj;
double rj[131072][3];
double mj[131072];
double ri[131072][3];
double a[131072][3];
double eps2 = 0.01;

int main()
{
  ..(A)..                              /* get initial condition for rj, mj, ri, a, and n */
  ..(B)..                              /* prepare result previously calculated by host */
  ..(C)..                              /* start stopwatch */
  force(rj, mj, ri, eps2, a, ni, nj);
  ..(D)..                              /* stop stopwatch and get the lap time */
  ..(E)..                              /* check results and calculate maximum cumulative error */
}
\end{verbatim}
\caption{A C program for measuring the performance and cumulative error of the CUNBODY-1. }
\label{fig:testcode}
\end{figure}

Figure.\ref{fig:testcode} shows our test program written by the standard C language to measure the
performance and cumulative error with the CUNBODY-1.
In this code, \verb|ri| and \verb|rj| represent positions of
$i$-particles and $j$-particles, respectively.
\verb|mj| represents mass of $j$-particles, \verb|a| represents force
calculated by a GPU thorough the CUNBODY-1 library, and \verb|eps2| is for
a softening parameter.  At the first step (A), the test program initializes
all values described above. At next step (B), the program
prepares results previously calculated by the host computer with
double precision.  In
(C), the time measurement begins and completes at (D).  Then the program
checks the errors for all forces calculated by the CUNBODY-1 library (\verb|force|). 
Note that, in the actual experiment, we execute the
\verb|force| function many times and average the times
for reducing the measurement deviation.

\subsection{System Details for Experiments}

Our library is implemented using the CUDA C language.  So we need the
CUDA library and also a GPU board and a GPU driver.  In
Table.\ref{tab:system}, we illustrate the details of our system for
the performance experiments.

\begin{table}
\caption{The system details for our experiment}
\begin{center}
\begin{tabular}{lll}
\hline
\hline
Component & & Comment \\
\hline
{\bf Hardware}  & &\\
GPU Board      & ASUS EN8800GTX/HTDP/768MByte       & \\
System Board   & ASUS P5WDG2 WS Professional        & \\
System Memory  & Transcend TS128MLQ64V8J            & DDR2-800 1 G Bytes \\
CPU            & Intel Core2Duo E6600               & 2.40 GHz\\
{\bf Software} & &\\
GPU driver     & NVIDIA Linux Display Driver Version 97.51 & {\tt http://www.nvidia.com/cuda/}\\
CUDA SDK       & CUDA SDK Version 0.8 for Linux x86 32-bit & {\tt http://www.nvidia.com/cuda/}\\
OS             & Debian GNU/Linux, kernel 2.6.18    & \\
Compiler       & gcc version 4.0.4                  & \\
\hline
\hline
\end{tabular}
\end{center}
\label{tab:system}
\end{table}

\subsection{Initial conditions}
We use the NEMO toolbox\cite{1995Teuben} to construct the
initial condition of a Plummer sphere with equal mass particles with
\verb|mkplummer| command. And we evaluate the cumulative errors between
force computed by the host computer with double precision and force
computed by the GPU.

\subsection{Cumulative Error}

\begin{table}
\caption{Results of the maximum cumulative errors for a Plummer sphere
with $N$ equal mass particles initially in virial equilibrium}
\Large
\begin{center}
\begin{tabular}{ccc}
\hline
\hline
$N$    & DIRECT style & Chamomile Scheme \\
\hline
2048   & 2.2$\times 10^{-6}$ & 5.4$\times 10^{\mathbf{-7}}$ \\
4096   & 3.0$\times 10^{-6}$ & 3.3$\times 10^{\mathbf{-7}}$ \\
8192   & 5.4$\times 10^{-6}$ & 5.0$\times 10^{\mathbf{-7}}$ \\
16384  & 7.0$\times 10^{-6}$ & 4.3$\times 10^{\mathbf{-7}}$ \\
32768  & 1.2$\times 10^{-5}$ & 6.8$\times 10^{\mathbf{-7}}$ \\
65536  & 1.4$\times 10^{-5}$ & 1.0$\times 10^{\mathbf{-6}}$ \\
131072 & 2.2$\times 10^{-5}$ & 1.5$\times 10^{\mathbf{-6}}$ \\
\hline
\hline
\end{tabular}
\end{center}
\label{tab:cumerror}
\end{table}

In Table.\ref{tab:cumerror}, we present the results of cumulative force
error with the initial conditions of a Plummer sphere with $N$ particles ($N=2048 \sim 131072$).  
The cumulative force errors listed mean the maximum force error for all of $N$ particles
, which is calculated by the following equation:
\begin{equation}
E_{\max} = \max_{i=0 \sim N-1}|\frac{|\mathbf{a}_{gpu,i} - \mathbf{a}_{host,i}|}{|\mathbf{a}_{host,i}|}|,
\label{eq:error_max}
\end{equation}
where $\mathbf{a}_{gpu,i}$ is force, which asserts on $i$-particle, calculated by the GPU.
$\mathbf{a}_{host,i}$ is
force on $i$-particle calculated by the host computer. 
$E_{\max}$ is the maximum error of the force.

The first column in Table.{\ref{tab:cumerror} represents the number of
particles, the second column is for the maximum error obtained by direct
accumulation of all $N$ $j$-particles in single precision(here after, the DIRECT
style accumulation). 
The last column is for the maximum error by the Chamomile Scheme.
Using the DIRECT style accumulation, the maximum cumulative error becomes
around $10^{-6}$.  On the other hand, our Chamomile Scheme is better
than the DIRECT style, in which the maximum cumulative error is suppressed around $10^{-7}$.

\subsection{Performance}

\begin{table}
\caption{Results of the performance measurements for a Plummer sphere
with $N$ equal mass particles initially in virial equilibrium, using
single GPU board and the CUNBODY-1 library}
\begin{center}
\LARGE
\begin{tabular}{ccc}
\hline
\hline
$N$ &  Computing time (sec) & Performance (Giga flop/s)\\
\hline
2048   &   9.21 $\times 10^{-4}$ & \bf{\Huge 173} \\
4096   &   2.99 $\times 10^{-3}$ & \bf{\Huge 213} \\
8192   &   1.08 $\times 10^{-2}$ & \bf{\Huge 235} \\
16384  &   4.14 $\times 10^{-2}$ & \bf{\Huge 246} \\
32768  &   0.162  & \bf{\Huge 251} \\
65536  &   0.642  & \bf{\Huge 254} \\
131072 &   2.548  & \bf{\Huge 256} \\
\hline
\hline
\end{tabular}
\end{center}
\label{tab:flops}
\end{table}


At the end of this section, we give our results of the measured
performances in Table.\ref{tab:flops}.  The first column represents
the number of particles, the second column is the time spent for the
CUNBODY-1 library (\verb|force|).  The last column is the performance
in Gflop/s(Giga flop/s).

We count flop/s by the following equation:
\begin{equation}
S  =  n_{flo} N^{2} / t  ~~[flop/s],
\label{eq:flops}
\end{equation}
where $N$ is the number of the particles.
$t$ is time spent for the CUNBODY-1 library.
$n_{flo}$ is the total number of the floating-point operations for
force calculation.
In this paper, we use 38 as $n_{flo}$.
This convection was introduced by Warren et al(1997)\cite{1997SC...Warren},
 and we follow it here since many researches use the same number
\cite{Makino_2003b}
\cite{2005astro-ph/0504407}
\cite{2006NewA...Nitadori}
\cite{2000PASJ...52..943H}
\cite{2005PASJ...inpress...H}.


\section{Discussion}

\subsection{Comparison of the GPU and the GRAPE in practical applications}

It seems to be almost clear that the performance of the GPU is
much better than that of the GRAPE in the measured performances.  
For the case of force calculation only, the theoretical peak performance of
GRAPE-6Af with four GRAPE-6 processor is only 87.5
Gflop/s\cite{2005astro-ph/0504407}. Oppositely, the GeForce8800GTX has
the peak performance of 518.4 Gflop/s, and our CUNBODY-1 with one
GeForce8800GTX has the measured performance of 173 $\sim$ 256 Gflop/s.

Of course, there remain many discussions in our insistence.  The main
discussion is about performance for other cases, such as the direct
integration with a Hermite Scheme.
In this case, we should calculate not
only force but also its time derivatives simultaneously.  For
this remained discussion, we have already got the measured performance for
force and its time derivatives calculation in simultaneous.  Our
current result has achieved 179 Gflop/s for 131072 particles with single
GeForce8800GTX\cite{Hamada_2007a}.

\subsection{Comparison with similar research projects}

\begin{figure}
\scriptsize
\begin{verbatim}
float4 force(float2 ij      : WPOS,
  uniform sampler2D pos)    : COLOR0
{
// Pos texture is 2D, not 1D, so we need to
// conver body index into 2D coords for pos tex
float4 iCoords  = getBodyCoords(ij);
float4 iPosMass = texture2D(pos, iCoords.xy);
float4 jPosMass = texture2D(pos, iCoords.zw);
float3 dir = iPos.xyz - jPos.xyz;
float r2 = dot(dir, dir);
dir = normalize(dir);
return dir * g * iPosMass.w * jPosMass.w / r2;
\end{verbatim}
\caption{Mark Harris's code}
\label{fig:HarrisCode}
\end{figure}

For $N$-body simulations, Nyland et
al.\cite{2004GPGPU..Nyland} and Harris et
al.\cite{2005Game..Harris}\cite{2005SIGGRAPH..Harris} reported
$N$-body simulations with GPUs.  Fig.\ref{fig:HarrisCode} shows a part
of Harris's implementation
\cite{2005Game..Harris}\cite{2005SIGGRAPH..Harris}
written in the Cg language.  This implementation is same as Nyland et
al\cite{2004GPGPU..Nyland}.  These implementations of force
calculation are based on the parallelism of particle pair, $(i,j)$.
They build an $N \times N$ force matrix (texture), $\mathbf{f}_{ij}$.
The float4 vector iPosMass contains three dimensional position vector
and mass of $i$-particle.  The total force acting
on $i$-particle is calculated by summing up with
$j$-particles by using binary tree summation.  This scheme is
performed very efficiently in parallel on a GPU, because it has
$N^2$-parallelism in the most heavy force calculation and
$N$-parallelism in relatively light calculation of summation. A minor
drawback of this algorithm is that the size of $\mathbf{f}_{ij}$ table
grows up in proportion to $O(N^2)$, and is limited to the maximum
texture size (2048 particle for their implementations). For more
particles, their implementations also need similar scheme to our
Chamomile Scheme, where the particles should be divided into blocks
and the total force should be calculated block by block.

%

\subsection{Performance modelling and more fine optimization}

There is a possibility that the performance will improve more for the
Chamomile Scheme.
When we almost completed this work, a new version of CUDA library was released
from the NVIDIA corporation.  The new version of CUDA library offers us
a method to access the texture memory spaces on the device memory.
The texture memory spaces are implemented as read only region of
the device memory and cached in on-chip memory spaces so as to improve the
effective bandwidth and latency. With the texture memory spaces, the
sampling costs one memory read from the device memory only on a cache
miss.
Current CUNBODY-1 library doesn't use any texture memory spaces.  So,
if we implement the I-POT on the texture memory spaces, the
performance might improve more. 
The improvement work needs to measure the number of clock cycles for
each internal operations. We can easily measure the clock cycles for
this purpose, because the CUDA library offers us a method to measure
the internal clock cycles. And we can construct the performance model
for the modified Chamomile Scheme\cite{Hamada_2007c}.

\subsection{Applying our scheme to the Hierarchical Tree Algorithm}

Hierarchical tree algorithm\cite{1986Natur.324..446B} is one of useful algorithms which
reduce the calculation cost from $O(N^2)$ to $O(N\log N)$.  Recently,
the hierarchical tree algorithm is widely used for investigating
extremely large-scale cosmological simulations with billions of
particles.  In such simulations with extreme numbers, we can
effectively use special-purpose hardwares for $O(N^2)$ interaction by
use of the modified tree algorithm developed by Barnes
(1990)\cite{B90}, firstly used on GRAPE-1A by Makino
(1990)\cite{Makino_1991}, and recently used on GRAPE-6Af by Fukushige
(2005)\cite{2005astro-ph/0504407}.

In the modified tree algorithm, not only the computing speed but also
the communication speed between the host computer and the special-purpose
hardware becomes very important, especially with the small number of
particles (around several thousands of particles).  In the modified tree
algorithm, the GPU seems to be quite suitable because
we can use quite high-speed link, the PCI-Express link. 
Using CUNBODY-1, the average transfer speed has got over 1.3 GByte/s with 2048 particles
\cite{Hamada_2007b} through the beta version of NVIDIA's
driver, currently limiting up to 2.0 GByte/s in maximum.

\subsection{Applying our scheme to other forms of interaction}

To tell the truth, our CUNBODY-1 library is implemented as a general
fashion to deal with arbitrary forms of interaction.  We have already
implemented libraries for other types of interaction not only for
gravitational force but also for smoothed particle hydrodynamics(SPH)
interactions, the time derivative of acceleration, gravitational
potential, van der Waals molecular interactions, boundary element
method (BEM) interactions, computer generated hologram(CGH)
interactions and Kolmogorov-Smirnov test interactions, working with
quite enough performance. We call such the generalized library as
CUNBODY(CUDA $N$-body)-1 kernel. Using CUNBODY-1 kernel, we can
implement a function which deals with arbitrary function form of
interaction just for users specifying the dimension of
particles($i$-particle, $j$-particle and the result of interactions)
etc, and its pipeline description (the CUNBODY pipeline description).
In appendix
\ref{cunbody-driver-pipe}, we present how we specify the dimension and
how we specify the pipeline description by CUNBODY-1 kernel\cite{Hamada_2007c}. 
We are planning to release the CUNBODY-1 kernel under the license of GPL(GNU
general Public License) soon.

\bigskip

This research was partially supported by the Special Postdoctoral
Research Program at RIKEN.
This research was partially supported by the Ministry of Education,
Culture, Sports, Science and Technology, Grant-in-Aid for Exploratory Research,
No.18654068, 2006.

\newpage
\appendix
\section{CUNBODY-1 user specified drivers and pipeline descriptions}
\label{cunbody-driver-pipe}
\begin{figure}
\scriptsize
\begin{verbatim}
#define IDIM  (4)
#define JDIM  (4)
#define FDIM  (3)

#include <cunbody_api.h>

static double p_xj[NMAX][JDIM];
static double p_xi[NMAX][IDIM];

void force(double xj[][3],
           double mj[],
           double xi[][3],
           double eps2,
           double a[][3],
           int ni,
           int nj)
{
  for(int i=0;i<ni;i++){
    for(int d=0;d<3;d++) p_xi[i][d] = xi[i][d];
    p_xi[i][3] = eps2;
  }
  for(int j=0;j<nj;j++){
    for(int d=0;d<3;d++) p_xj[j][d] = xj[j][d];
    p_xj[j][3] = mj[j];
  }

  cunbody_api(p_xj, p_xi, a, ni, nj);

}
\end{verbatim}
\caption{A CUNBODY-1 driver code for force function.}
\label{fig:cunbody-force-driver}
\end{figure}

\begin{figure}
\scriptsize
\begin{verbatim}
#define IDIM  (6)
#define JDIM  (7)
#define FDIM  (7)

#include <cunbody_api.h>

static double p_xj[NMAX][JDIM];
static double p_xi[NMAX][IDIM];
static double p_fi[NMAX][FDIM];

void force(double x[][3],
           double v[][3],
           double m[],
           double p[],
           double a[][3],
           double jk[][3],
           int n)
{
  for(int i=0;i<n;i++){
    for(int d=0;d<3;d++) {
      p_xi[i][d]   = x[i][d];
      p_xi[i][d+3] = v[i][d];
      p_xj[i][d]   = x[i][d];
      p_xj[i][d+3] = v[i][d];
    }
    p_xj[i][6] = m[i];
  }

  cunbody_api(p_xj, p_xi, p_fi, n, n);

  for(int i=0;i<n;i++){
    for(int d=0;d<3;d++){
      a[i][d]  = p_fi[i][d];
      jk[i][d] = p_fi[i][d+3];
    }
    p[i] = p_fi[i][6];
  }
}
\end{verbatim}
\caption{A CUNBODY-1 code for force, jerk and potential function.}
\label{fig:cunbody-forcejerk-driver}
\end{figure}

\begin{figure}
\scriptsize
\begin{verbatim}
float xi,yi,zi,eps2;
float xj,yj,zj,mj;
float dx,dy,dz;
float dx2,dy2,dz2;
float r2,r1i,r2i,r3i;
float mr3i;

xi   = INPUT_I[0];
yi   = INPUT_I[1];
zi   = INPUT_I[2];
eps2 = INPUT_I[3];

xj   = INPUT_J[0];
yj   = INPUT_J[1];
zj   = INPUT_J[2];
mj   = INPUT_J[3];

dx = xj - xi;
dy = yj - yi;
dz = zj - zi;
dx2 = dx * dx;
dy2 = dy * dy;
dz2 = dz * dz;
r2 = (dx2 + dy2) + (dz2 + eps2);
r1i = 1/sqrt(r2);
r2i = r1i * r1i;
r3i = r2i * r1i;
mr3i = mj * r3i;

OUTPUT[0] += mr3i * dx;
OUTPUT[1] += mr3i * dy;
OUTPUT[2] += mr3i * dz;
\end{verbatim}
\caption{A CUNBODY-1 pipeline description for force interaction.}
\label{fig:cunbody-force-pipe}
\end{figure}

\begin{figure}
\scriptsize
\begin{verbatim}
float eps2 = 1.0/(256.0*256.0);
float xj_0, xj_1, xj_2, xi_0, xi_1, xi_2, dx_0, dx_1, dx_2, dx2_0, dx2_1, dx2_2;
float vj_0, vj_1, vj_2, vi_0, vi_1, vi_2, dv_0, dv_1, dv_2, xv_0, xv_1, xv_2;
float r2, r1i, r2i, r3i, r5i, mj, mf, mr5i, mf3a, fx_0, fx_1, fx_2, pot;
float jk_0,  jk_1,  jk_2, jk1_0, jk1_1, jk1_2, jk2_0, jk2_1, jk2_2;

xi_0 = INPUT_I[0];
xi_1 = INPUT_I[1];
xi_2 = INPUT_I[2];
vi_0 = INPUT_I[3];
vi_1 = INPUT_I[4];
vi_2 = INPUT_I[5];

xj_0 = INPUT_J[0];
xj_1 = INPUT_J[1];
xj_2 = INPUT_J[2];
vj_0 = INPUT_J[3];
vj_1 = INPUT_J[4];
vj_2 = INPUT_J[5];
mj   = INPUT_J[6];

dx_0 = xj_0 - xi_0;
dx_1 = xj_1 - xi_1;
dx_2 = xj_2 - xi_2;
dx2_0 = dx_0 * dx_0;
dx2_1 = dx_1 * dx_1;
dx2_2 = dx_2 * dx_2;
r2   = dx2_0 + dx2_1 + dx2_2 + eps2;
r1i  = 1/sqrt(r2);
r2i  = r1i * r1i;
r3i  = r1i * r2i;
r5i  = r2i * r3i;
mr5i = mj * r5i;
mf   = mr5i * r2;
pot  = mf * r2;
fx_0 = mf * dx_0;
fx_1 = mf * dx_1;
fx_2 = mf * dx_2;
dv_0   = vj_0 - vi_0;
dv_1   = vj_1 - vi_1;
dv_2   = vj_2 - vi_2;
jk1_0  = mf * dv_0;
jk1_1  = mf * dv_1;
jk1_2  = mf * dv_2;
xv_0   = dx_0 * dv_0;
xv_1   = dx_1 * dv_1;
xv_2   = dx_2 * dv_2;
mf3a   = 3.0 * mr5i * (xv_0 + xv_1 + xv_2);
jk2_0  = mf3a * dx_0;
jk2_1  = mf3a * dx_1;
jk2_2  = mf3a * dx_2;
jk_0   = jk1_0 - jk2_0;
jk_1   = jk1_1 - jk2_1;
jk_2   = jk1_2 - jk2_2;

OUTPUT[0] += fx_0;       // acc_x
OUTPUT[1] += fx_1;       // acc_y
OUTPUT[2] += fx_2;       // acc_z
OUTPUT[3] += jk_0;       // jerk_x
OUTPUT[4] += jk_1;       // jerk_y
OUTPUT[5] += jk_2;       // jerk_z
OUTPUT[6] -= pot;        // pot
\end{verbatim}
\caption{A CUNBODY-1 pipeline description for force, jerk and potential interaction.}
\label{fig:cunbody-forcejerk-pipe}
\end{figure}

\end{document}